\documentclass[preprint,aps,showpacs]{revtex4}
 \usepackage{amssymb} \usepackage{graphicx}

\begin{document}
 \title{Killing-Yano Forms of a Class of Spherically Symmetric Space-Times II:
  A Unified Generation of Higher Forms}
 \author{\"{O}. A\c{c}{\i}k $^{1}$}
 \email{ozacik@science.ankara.edu.tr}
 \author{\"{U}. Ertem $^{1}$}
 \email{uertem@science.ankara.edu.tr}
\author{M. \"{O}nder $^{2}$}
 \email{onder@hacettepe.edu.tr}
 \author{A. Ver\c{c}in $^{1}$}
 \email{vercin@science.ankara.edu.tr}
\address{$^{1}$ Department of Physics, Ankara University, Faculty of Sciences,
06100, Tando\u gan-Ankara, Turkey\\
$^{2}$ Department of Physics Engineering, Hacettepe University,
06800, Beytepe-Ankara, Turkey.}

\date{\today}

\begin{abstract}
Killing-Yano (KY) two and three forms of a class of spherically
symmetric space-times that includes the well-known  Minkowski,
Schwarzschild, Reissner-Nordstr{\o}m, Robertson-Walker and six
different forms of  de Sitter space-times as special cases are
derived in a unified and exhaustive manner. It is directly proved
that while the Schwarzschild and Reissner-Nordstr{\o}m space-times
do not accept any KY 3-form and they accept only one 2-form, the
Robertson-Walker space-time admits four KY 2-forms and only one KY
3-form. Maximal number of KY-forms are obtained for Minkowski and
all known forms of de Sitter space-times. Complete lists comprising
explicit expressions of KY-forms are given.
\end{abstract}
 \maketitle

\section{Introduction}

In the previous study \cite{ozumav1} (henceforth referred to as I),
we developed a constructive method which provided a unified
generation of all Killing vector fields for a class of four
dimensional ($4D$) spherically symmetric space-times. As the sequel
of I, in the present paper we shall investigate the explicit forms
of KY two and three forms  by solving the KY-equations
\begin{eqnarray}
\nabla_{X_{a}}\omega_{(p)} =\frac{1}{p+1} i_{X_{a}}d\omega_{(p)}\;,
\end{eqnarray}
for this class of space-time metrics in the $p=2$ and $p=3$ cases.

The underlying base manifold is supposed to be a $4D$
pseudo-Riemannian manifold endowed with the metric $g$ having the
Lorentzian signature $(-+++)$ such that
\begin{eqnarray}
g=-e^{0}\otimes e^{0}+e^{1}\otimes e^{1}+e^{2}\otimes
e^{2}+e^{3}\otimes e^{3} \;.\nonumber
\end{eqnarray}
in a local orthonormal co-frame basis $\{e^{a}\}$. This metric can
be parameterized by the (metric) characterizing functions
$T=\exp(\lambda(t))$ and $H_{j}=H_{j}(r)$ by choosing
\begin{eqnarray}
e^0 &=& H_{0}dt\;,\qquad
e^1=T H_{1}dr\;,\nonumber\\
e^2 &=& T H_{2}d\theta\;,\quad e^3=T H_{2}\sin\theta d\varphi
\;.\nonumber
\end{eqnarray}
The corresponding orthonormal vector bases will be denoted by
$X_{a}$. The torsion-free connection $1$-forms for this class of
metrics and covariant derivatives of basis elements required for
explicit calculations can be found in I. We shall mainly use the
notation of I, which is in accordance with \cite{Benn-Tucker}.

As they span the kernel of the linear operator
$\nabla_{X}-(p+1)^{-1}i_{X}d$, the space $Y_{p}$ of all KY p-forms
constitute a linear space \cite{Semmelman1,Stepanov}. Any function
is a KY 0-form, $\omega_{(1)}$ is the dual of a Killing vector field
and $\omega_{(n)}$ is a constant (parallel), that is, it is a
constant multiple $\omega_{(n)}=az$ of the volume form, say
$z=e^{0123}$ for $n=4$. Therefore while $Y_{0}$ is infinite
dimensional, $Y_{n}$ is one dimensional. By a straightforward
extension of the argument for determining the maximum number of
Killing vectors, the upper bound for the dimension of $Y_p$'s can
now be established for each $p$ \cite{Kastor,Kastor1}. For this
purpose, we should first note that equation (1) suggests an
equivalent definition: a p-form is a KY p-form if and only if its
symmetrized covariant derivatives vanish, that is, if and only if
\begin{eqnarray}
i_{X}\nabla_{Y}\omega_{(p)}+
i_{Y}\nabla_{X}\omega_{(p)}=0\;,\nonumber
\end{eqnarray}
is satisfied for all pair of the vector fields $X$ and $Y$. This is
also equivalent to $i_{X}\nabla_{X}\omega_{(p)}=0$ and in particular
every KY-form is divergent free, or equivalently co-closed, that is,
$\delta \omega_{(p)}=-i_{X^{a}}\nabla_{X_{a}}\omega_{(p)}=0$. These
statements imply that the covariant derivatives of KY forms are also
totally anti-symmetric, with respect to the additional tensorial
index. Hence the maximum numbers of linearly independent components
and their first covariant derivatives are $C(n,p)$ and $C(n,p+1)$
respectively, where $C(n,p)$ stands for the binomial numbers.

On the other hand ``the second covariant derivatives'', that is, the
Hessian of KY forms can be written, in terms of the curvature
2-forms $R^{c}_{\;a}$ as
\begin{eqnarray}
(\nabla_{X_{a}}\nabla_{X_{b}}-\nabla_{\nabla_{X_{a}}X_{b}})\omega_{(p)}
=\frac{1}{p} i_{X_{b}}(R^{c}_{\;a}\wedge
i_{X_{c}}\omega_{(p)})\;.\nonumber
\end{eqnarray}
These imply that the value of any component of a KY-form at any
point is entirely determined by the value of its first covariant
derivative and by the value of the component itself at the same
point. As a result the upper bounds for the numbers of linearly
independent KY-forms is determined by the sum
$C(n,p)+C(n,p+1)=C(n+1,p+1)$. In particular for $n=4$ the upper
bounds are $10,10$ and $5$, respectively, for $p=1, p=2$ and $p=3$.
These bounds are attained for the space-times of constant curvature.

The rest of the paper is structured in two main parts. In the next
section, the defining equations for KY 2-forms and their
integrability conditions are obtained. By the integrability
conditions, the set of solutions naturally breaks up into cases and
subcases that are considered in the rest of first part consisting
five sections. The second part is entirely devoted to determination
of KY 3-forms. The final Section VIII presents a summary of the
results, and we briefly point out how to calculate the associated
Killing tensors and related linear and quadratic first integrals for
the considered class of space-times.

\section{KY 2-Forms: Defining Equations, Integrability Conditions}

For a 2-form
\begin{eqnarray}
\omega_{(2)}=\alpha e^{01}+\beta e^{02}+\gamma e^{03}+\delta
e^{12}+\epsilon e^{13}+\mu e^{23} \;,\nonumber
\end{eqnarray}
there are two types of equations that result from KY-equation (1) in
the case of $p=2$. The first consists of twelve relatively simple
equations resulting from the left hand side of the KY equations. Two
of them are $\alpha_t=0=\alpha_r$, which imply that
$\alpha=\alpha(\theta,\varphi)$, and the other ten equations are as
follows:
\begin{eqnarray}
\beta_{t}&=&\frac{H'_0}{T H_1}\delta \;,\;\qquad \qquad
\beta_{\theta}=-\frac{H'_2}{H_1}\alpha\;,\nonumber\\
\delta_{r} &=& \dot{T}\frac{H_1}{H_0} \beta\;,\;\qquad \qquad
\delta_{\theta} =-\dot{T}\frac{H_2}{H_0}\alpha\;,\nonumber\\
\gamma_{t} &=&\frac{H'_0}{T H_1}\epsilon \;,\;\qquad \qquad
\gamma_{\varphi}=
-\frac{H'_2}{H_1}\sin\theta \alpha -\cos\theta\beta\;, \\
\epsilon_{r} &=& \dot{T}\frac{H_1}{H_0}\gamma\;,\;\qquad \qquad
\epsilon_{\varphi} =
-\dot{T}\frac{H_2}{H_0}\sin\theta \alpha -\cos\theta\delta\;,\nonumber\\
\mu_{\theta} &=&\dot{T} \frac{H_2}{H_0}\gamma
-\frac{H'_2}{H_1}\epsilon  \;, \quad \mu_{\varphi} =-\dot{T}
\frac{H_2}{H_0} \sin\theta\beta +
\frac{H'_2}{H_1}\sin\theta\delta\;.\nonumber
\end{eqnarray}
The second type also consists of twelve equations, but four of them
are just a sum of two other equations of the same type. Hence, there
are eight independent additional equations, six of which can be put
in the following forms:
\begin{eqnarray}
\alpha_{\theta}
&=&-\frac{H^{2}_{2}}{H_1}\partial_{r}\frac{\beta}{H_{2}}\;,\quad
\alpha_{\varphi}=- \frac{H_{2}^{2}}{H_1}\sin
\theta\partial_{r}\frac{\gamma}{H_{2}}\;,\nonumber\\
\partial_{r}\frac{\beta}{H_{0}}&=&-T^{2}\frac{H_{1}}{H^{2}_{0}}
\partial_{t}\frac{\delta}{T}\;,\quad
\gamma_{\theta} =-T^{2}\frac{H_{2}}{H_{0}}\partial_{t}\frac{\mu}
{T}\;,\\
\partial_{r}\frac{\gamma}{H_{0}}&=&-
T^{2}\frac{H_{1}}{H^{2}_{0}}\partial_{t}\frac{\epsilon}{T}\;,\quad
\epsilon_{\theta} =-\frac{H^{2}_{2}}{H_{1}}\partial_{r}\frac{\mu}
{H_{2}}\;.\nonumber
\end{eqnarray}
The last two equations (presented below) do not depend on the metric
coefficient functions, and therefore much of the essence of
spherical symmetry is encoded in them:
\begin{eqnarray}
\beta_{\varphi}
&=&-\sin^{2}\theta\partial_{\theta}\frac{\gamma}{\sin\theta}\;,
\quad\quad \delta_{\varphi}=-
\sin^{2}\theta\partial_{\theta}\frac{\epsilon}{\sin\theta}\;.
\end{eqnarray}

Before solving these eighteen equations, there are some
integrability conditions that must be met. Two of them, which enable
us to find the solutions in a systematic way, are
\begin{eqnarray}
\dot{T}H^{\prime}_{0}\alpha &=& 0\;,\\
\;[\rho(r)+\varrho(t)\;]\alpha &=& 0\;,
\end{eqnarray}
where
\begin{eqnarray}
\rho=\frac{H_{0}^{3}}{H_{1}H_{2}}(\frac{H_{2}^{\prime}}{H_{0}H_{1}})^{\prime}\;,\quad
\varrho=T\ddot{T}-\dot{T}^{2}=T^{2}\partial_t(\frac{\dot{T}}{T})\;.
\end{eqnarray}
Condition (5) separately follows from each of
\begin{eqnarray}
\beta_{t\theta}=\beta_{\theta t},\;\gamma_{t\varphi}=\gamma_{\varphi
t},\;\delta_{r\theta}=\delta_{\theta
r}\;,\;\epsilon_{r\varphi}=\epsilon_{\varphi r} \;,\nonumber
\end{eqnarray}
and condition (6) can be checked from
$\delta_{t\theta}=\delta_{\theta t}$ or $\epsilon_{\varphi
t}=\epsilon_{t\varphi}$. The integrability conditions imply that
solutions must be investigated in two classes: (A) $\alpha\neq 0$
and (B) $\alpha=0$, such that the first consists of three important
subclasses characterized by
\begin{eqnarray}
(i)\;H_0^{\prime}=0\;,\;\;\dot{T}\neq 0\;,\quad (ii)\;
H_0^{\prime}\neq0\;,\;\;\dot{T}=0\;,\quad (iii)\;
H_0^{\prime}=0=\dot{T}\;.\nonumber
\end{eqnarray}
In fact there are $6\times 6$ integrability conditions, some of
which are trivially satisfied and the remaining ones will be
considered in the following classes. We should also note that
whenever $T$ or $H_{k}$'s are constant, they must be considered to
be nonzero to keep the metric non-degenerate.

In the next three sections, we shall consider the first (A) case in
which $\alpha$ is different from zero. At the outset of these
sections, we should note the following two equations for $\alpha$:
\begin{eqnarray}
\alpha_{\theta\theta}+\ell_{1}\alpha &=&0\;,\\
\alpha_{\varphi\varphi}+\ell_{1}\sin^{2}\theta
\alpha+\sin\theta\cos\theta\alpha_{\theta}&=&0\;,
\end{eqnarray}
which do not change in the following subcases. These are obtained by
differentiating  the first two equations of (3) and by making use of
the equation for $\beta_\theta$ and $\gamma_\varphi$ from (2). Here
the constant $\ell_{1}$ is defined, in terms of
$P=H_{2}^{\prime}/H_{1}H_2$, as follows:
\begin{eqnarray}
\ell_{1}=-\frac{H_{2}^{2}}{H_{1}}P^{\prime}\;,
\end{eqnarray}

We should finally note that, as there are no equations involving the
first power of $\mu$,
\begin{eqnarray}
\omega_{0}=TH_2e^{23} \;\nonumber
\end{eqnarray}
is a solution of the KY-equation without any constraint. This
observation means that all the space-times within the considered
class of metrics admit of at least one KY 2-form. This fact was also
observed for the static space-times in \cite{Howarth,Collinson 1}.
In fact, it turns out that $\omega_{0}$ is the only KY 2-form
admitted by two important cases, the Schwarzschild and the
Reissner-Nordstr{\o}m space-times, which do not accept any KY
3-forms.

\section{KY 2-Forms for $H_0^{\prime}=0,\;\dot{T}\neq0\;,\alpha\neq 0$}

In the case of nonzero $\alpha$ and $H_0^{\prime}=0$, two equations
of (2) show that $\beta$ and $\gamma$ are time independent and the
integrability conditions (5-7) require that $\ell$, defined in terms
of $P$ by
\begin{eqnarray}
\frac{H_{0}^{2}}{H_{1}H_{2}}(H_{2}P)^{\prime}=-\ell=T^{2}\partial_t(\frac{\dot{T}}{T})\;,
\end{eqnarray}
must be a constant. By multiplying both sides of the first equality
by $H_2H_{2}^{\prime}$ it can easily be integrated to write
\begin{eqnarray}
P^{2}=\frac{\ell_1}{H^{2}_2}-\frac{\ell}{H^{2}_0}\;,
\end{eqnarray}
where $\ell_1$ is taken as an integration constant, since the
defining relation (10) of $\ell_1$ is implied by the condition (12).

Then, in terms of
\begin{eqnarray}
{\cal B}&=&\dot{T}\frac{H_2}{H_0}\beta\;,\quad
{\cal D}=H_2P\delta\;,\nonumber\\
{\cal G}&=&\dot{T}\frac{H_2}{H_0}\gamma\;,\quad {\cal
E}=H_2P\epsilon\;,\nonumber
\end{eqnarray}
we define
\begin{eqnarray}
x={\cal G}-{\cal E}\;,\quad y={\cal B}-{\cal D}\;.
\end{eqnarray}
The first four equations appearing in the second column of (2) and
two equations of (4) give the following equations for $x$ and $y$:
\begin{eqnarray}
y_\theta=0\;,\quad x_\varphi=-\cos\theta y\;,\quad
y_\varphi=-\sin^{2}\theta\partial_\theta\frac{x}{\sin\theta}\;.
\end{eqnarray}
These immediately imply $y_{\varphi\varphi}+y=0$ and hence
\begin{eqnarray}
y=y_1\cos\varphi+y_2\sin\varphi\;,\quad x=\cos\theta
y_\varphi+y_3\sin\theta\;,
\end{eqnarray}
where three $y_i$ functions depend on both $t$ and $r$, and the
expression of $x$ is obtained by integration from (14).

\subsubsection{The general forms of $\gamma,\epsilon$ and $\mu$}

In terms of $x$ and $y$, the last two equations of (2) are simply
$\mu_\theta=x,\;\mu_\varphi=-y\sin\theta$, and therefore they
specify the general form of $\mu$ as
\begin{eqnarray}
\mu=y_\varphi\sin\theta-y_3\cos\theta+c_0TH_2\;,
\end{eqnarray}
where $c_0$ is an integration constant. In specifying the last term
of (16), we make use of the $\gamma_\theta$ and
$\epsilon_\theta$-equations of (3), which transform to
\begin{eqnarray}
\gamma_\theta=T^{2}\frac{H_2}{H_0}\partial_t\frac{x_\theta}{T}\;,\quad
\epsilon_\theta=\frac{H_2^{2}}{H_1}\partial_r\frac{x_\theta}{H_2}\;.
\end{eqnarray}
Integration of these equations with respect to $\theta$ yield
\begin{eqnarray}
\gamma=T^{2}\frac{H_2}{H_0}\partial_t\frac{x}{T}+g(r,\varphi)\;,\quad
\epsilon=\frac{H_2^{2}}{H_1}\partial_r\frac{x}{H_2}+\varepsilon(t,r,\varphi)\;.
\end{eqnarray}
By substituting these expression into $x={\cal G}-{\cal E}$, we
obtain
\begin{eqnarray}
\varepsilon=\frac{\dot{T}}{H_0}\frac{g}{P}\;,
\end{eqnarray}
and a set of three equations for $y_i$, which in terms of
$Y_i=y_i/TH_2$ can be written as
\begin{eqnarray}
Y_i=T\dot{T}(\frac{H_2}{H_0})^{2}Y_{it}-\frac{H_2^{2}P}{H_1}
Y_{ir}\;, \quad i=1,2,3\;.
\end{eqnarray}
In writing equations (20), we have equated the coefficient functions
of different trigonometric functions forming a basis, and relation
(19) results from the fact that $x$ does not contain a term
independent from $\theta$.

We now concentrate on specifying $g$ and $y_i$ functions. For this
purpose we consider the coupled $\epsilon_r$-equation of (2) and
$\gamma_r$-equation of (3) which, in terms of $P$ and $\ell$, give
\begin{eqnarray}
\partial_r(\frac{g}{P})=H_1g\;,\quad
g_r=-\frac{\ell}{H_0^{2}}\frac{H_1}{P}g\;,
\end{eqnarray}
in addition to two sets of equations for $y_i$ :
\begin{eqnarray}
\partial_r(\frac{H^{2}_2}{H_1}Y_{ir})&=&T\dot{T}\frac{H_1H^{2}_2}{H^{2}_0}Y_{it}\;,\\
\partial_r(H^{2}_2Y_{it})&=&-H^{2}_2Y_{itr}\;.
\end{eqnarray}
By dividing the first equation of (21) with $P$, it can be easily
integrated to yield
\begin{eqnarray}
g=H_2Pu_{\varphi}\;,
\end{eqnarray}
where for convenience the last factor of $g$ has been written as the
derivative of $u=u(\varphi)$. From the second equation of (21) we
then obtain just one of the conditions of (11), when
$u_{\varphi}\neq 0$.

After a slight rearrangement, the equation (23) can be integrated to
find
\begin{eqnarray}
Y_{i}=\frac{q_i(t)}{H_2}+\frac{z_i(r)}{H_2}\;
\end{eqnarray}
The substitution of this relation into (20) and (22) yield
\begin{eqnarray}
T\dot{T}\dot{q}_{i}-[\ell+(1-\ell_1)\frac{H^{2}_0}{H_2^{2}}]
q_i&=&[\ell+(1-\ell_1)\frac{H_{0}^{2}}{H_1}]z_i+H_{0}^{2}\frac{P}{H_1}z_{i}^{\prime}\;,\\
T\dot{T}\dot{q}_{i}-\ell q_i&=&\ell
z_i-\frac{H_{0}^{2}H_1^{\prime}}{H_1^{3}}z_i^{\prime}+\frac{H_0^{2}}{H^{2}_1}z_i^{\prime\prime}
\;,
\end{eqnarray}
for $i=1,2,3$. As we are about to see at the beginning of next
section, the constant $\ell_1$ must be $1$. Anticipating this result
here, we see that the above equations are separable, and therefore
each side of them must be a constant such that
\begin{eqnarray}
T\dot{T}\dot{q}_{i}-\ell q_i&=&m_i\;,\nonumber\\
\ell
z_i-\frac{H_{0}^{2}H_1^{\prime}}{H_1^{3}}z_i^{\prime}+\frac{H_0^{2}}{H^{2}_1}z_i^{\prime\prime}
&=&m_i\;,\\
\ell z_i+\frac{H_{0}^{2}}{H_1}Pz_{i}^{\prime}&=&m_i\;.\nonumber
\end{eqnarray}
It is not hard to see that the last two sets of (28) can be
integrated to obtain
\begin{eqnarray}
z_i=c_{i}H_{2}P-\tilde{c}_{i}\;,\quad i=1,2,3
\end{eqnarray}
with $m_i=-\ell \tilde{c}_{i}$. The equations in the first line of
(28) need not be integrated at this point, since they will be
directly solved during the investigation of next subsection.

\subsubsection{The general forms of $\alpha,\beta$ and $\delta$}

Relations found by (25) specify $y_i$ for $i=1,2,3$ as follows :
\begin{eqnarray}
y_{i}=Tq_i(t)+Tz_i(r)\;.
\end{eqnarray}
By noting that $(y_{i}/T)_{tr}=0$, the substitution of the
expression of $\gamma$ given by (18) into the second equation of (3)
yield $\alpha_\varphi=\ell_1\sin\theta u_\varphi$ in view of (24).
This can easily be integrated to write $\alpha=\ell_1\sin\theta
u+f(\theta)$, and then by using this in equations (8) and (9) we
obtain
\begin{eqnarray}
\ell_1(\ell_1-1)\sin\theta u+f_{\theta\theta}+\ell_{1}f &=&0\;,\nonumber\\
\ell_1[u_{\varphi\varphi}+(\cos^{2}\theta+\ell_{1}\sin^{2}\theta)
u]+\ell_1\sin\theta f+\cos\theta f_{\theta}&=&0\;.\nonumber
\end{eqnarray}
Since $u$ is a functions of $\varphi$, these relations imply that
$\ell_1$ must be either $0$ or $1$. In fact, the former value is not
compatible with the  very definition of spherical symmetry, since it
would lead to metric coefficient functions depending on some finite
powers of angular coordinates. Indeed, equations (8) and (9) show
that when $\ell_1=0$, $\alpha$ can be assumed to be a nonzero
constant. But the equations in the second column of (2) then show
that there is no way to get rid of the above mentioned angular
dependence. Therefore from now on, we take $\ell_1=1$ by which the
above two equations are reduced to the following forms :
\begin{eqnarray}
f_{\theta\theta}+f=0\;,\quad u_{\varphi\varphi}+u+\sin\theta
f+\cos\theta f_{\theta}=0\;.
\end{eqnarray}
Hence, in terms of integration constants $c_4,c_5,c_6, \tilde{c}_4$
and
\begin{eqnarray}
v=v(\varphi)=c_5\cos\varphi+c_6\sin\varphi\;,
\end{eqnarray}
we have $f=c_4\cos\theta+\tilde{c}_4\sin\theta$ and
$u=v-\tilde{c}_4$ which implies $u_\varphi=v_\varphi$. These
completely specify $\alpha$ as
\begin{eqnarray}
\alpha=c_4\cos\theta+\sin\theta v\;.
\end{eqnarray}

Using (33) in the second equation of (2) and the first equation of
(3) leads us to
\begin{eqnarray}
\beta=H_2P\alpha_\theta+H_2\bar{f}(\varphi)\;,\quad \delta=\dot{T}
\frac{H_2}{H_{0}}(\alpha_\theta+\frac{\bar{f}}{P})-\frac{y}{H_2P}
\;,
\end{eqnarray}
where $\delta$ is obtained from the definition $y={\cal B}-{\cal
D}$. Substitutions of the solutions (33) and (34) into the
$\gamma_\varphi$-equation of (2) yield, in terms of constants $c_7$
and $c_8$
\begin{eqnarray}
\bar{f}=\frac{v_1}{H_{0}}\;,\quad v_1=v_1(\varphi)=c_7\cos\varphi
+c_8\sin\varphi\;,
\end{eqnarray}
and $\dot{q}_{1}=c_7/T^{2},\;\dot{q}_{2}=c_8/T^{2}$ since $\gamma$
is independent of time.

\subsubsection{KY 2-forms for time-dependent de Sitter space-time: $\ell\ne 0$ solutions}

From the first set of equations of (28), $q_1$ and $q_2$ can be
completely specified as
\begin{eqnarray}
q_{1}=c_7\frac{\dot{T}}{\ell T}+\tilde{c}_1\;,\quad
q_{2}=c_8\frac{\dot{T}}{\ell T}+\tilde{c}_2\;,
\end{eqnarray}
by recalling the relations $m_1=-\ell \tilde{c}_1$ and $m_2=-\ell
\tilde{c}_2$. In view of (29), (30) and (36), $y$ is also completely
specified as
\begin{eqnarray}
y=\frac{\dot{T}}{\ell}v_1+TH_2Pv_2\;,
\end{eqnarray}
where we have defined
\begin{eqnarray}
v_2=v_2(\varphi)=c_1\cos\varphi +c_2\sin\varphi\;.
\end{eqnarray}

In view of (35) and (37), while $\beta$ and $\delta$ have been
completely determined, there remains to determine $q_3(t)$ of the
$y_3$-function
\begin{eqnarray}
y_{3}=Tq_3(t)+T(c_{3}H_{2}P-\tilde{c}_{3})\;
\end{eqnarray}
for complete specification of $\gamma,\epsilon$ and $\mu$. The
resulting $\beta,\;\delta$ solutions identically satisfy the
$\beta_r$-equation of (3) and $\delta_r$-equation of (2), and
therefore they give nothing new. On the other hand, if (37) and (38)
are used in the $\gamma$-expression of (18), we see that like $q_1$
and $q_2$, $q_3$ must be equal to $c_9(\dot{T}/\ell T)+\tilde{c}_3$
since $\gamma$ does not depend on $t$. The resulting $\gamma$ and
$\epsilon$ also satisfy the $\epsilon_r$-equation of (2), as well as
the definition $x={\cal G}-{\cal E}$.  But, as they do not take part
in any metric coefficient functions, the constants
$\tilde{c}_1,\tilde{c}_2$ and $\tilde{c}_3$ become redundant. We are
now ready to present all the coefficient functions together :
\begin{eqnarray}
\alpha&=&c_4\cos\theta+\sin\theta v\;,\quad
\beta=H_2P\alpha_\theta+\frac{H_2}{H_0}v_1\;,\nonumber\\
\gamma&=& \frac{H_2}{H_{0}}(c_9\sin\theta+\cos\theta
v_{1\varphi})+H_2Pv_\varphi\;,\nonumber\\
\delta &=& \dot{T}
\frac{H_2}{H_{0}}\alpha_\theta-\frac{1}{\ell}\dot{T}H_2Pv_1-Tv_2 \;,
\\
\epsilon&=&-\sin\theta(c_9\frac{1}{\ell}\dot{T}H_2 P+
c_3T)-\cos\theta(\frac{1}{\ell}\dot{T}H_2Pv_{1\varphi}+Tv_{2\varphi})+\dot{T}\frac{H_2}{H_0}
v_\varphi\;,\nonumber\\
\mu&=&\sin\theta(\frac{1}{\ell}\dot{T}v_{1\varphi}+
TH_2Pv_{2\varphi})-\cos\theta(\frac{c_9}{\ell}\dot{T}+c_3TH_2P)+c_0TH_2\;,\nonumber
\end{eqnarray}
which determine ten linearly independent KY $2$-forms, one for each
$c_i,\;i=0,1,\ldots,9$:
\begin{eqnarray}
\omega_{0}&=&TH_2e^{23}\;,\nonumber\\
\omega_{1}&=&-T\cos\varphi e^{12}+T\sin\varphi(\cos\theta e^{13}+H_2P\sin\theta e^{23})\;,\nonumber\\
\omega_{2}&=& -T\sin\varphi e^{12}-T\cos\varphi(\cos\theta e^{13}-H_2P\sin\theta e^{23})\;,\nonumber\\
\omega_{3}&=& -T(\sin\theta e^{13}+H_{2}P\cos\theta e^{23})
\;,\nonumber\\
\omega_{4}&=& \cos\theta e^{01}-H_2\sin\theta\Omega_1\wedge e^{2}\;,\\
\omega_{5}&=& \cos\varphi(\sin\theta
e^{01}+H_2\cos\theta\Omega_1\wedge e^{2})
-H_2\sin\varphi\Omega_1\wedge e^{3}\;,\nonumber\\
\omega_{6}&=& \sin\varphi(\sin\theta
e^{01}+H_2\cos\theta\Omega_1\wedge e^{2})
+H_2\cos\varphi\Omega_1\wedge e^{3}\;,\nonumber\\
\omega_{7}&=&H_2\cos\varphi \Omega_2\wedge
e^{2}-H_2\sin\varphi(\cos\theta\Omega_2\wedge
e^{3}+\frac{\dot{T}}{\ell}\sin\theta e^{23})\;,\nonumber\\
\omega_{8}&=&H_2\sin\varphi \Omega_2\wedge e^{2}+H_2\cos\varphi
(\cos\theta\Omega_2\wedge
e^{3}+\frac{\dot{T}}{\ell}\sin\theta e^{23})\;,\nonumber\\
\omega_{9}&=&H_2\sin\theta \Omega_2\wedge
e^{3}-\frac{\dot{T}}{\ell}\cos\theta e^{23} \;.\nonumber
\end{eqnarray}
where the 1-forms $\Omega_1$ and $\Omega_2$  are defined, for the
sake of simplicity, as
\begin{eqnarray}
\Omega_1=Pe^{0}+\frac{\dot{T}}{H_0}e^{1}\;,\quad
\Omega_2=\frac{1}{H_0}e^{0}-\frac{\dot{T}}{\ell}Pe^{1}\;.
\end{eqnarray}

\subsubsection{KY 2-forms for the usual form of de Sitter space-time: $\ell=0$ solutions}

When $\ell$ is zero, the integrability conditions (11) and condition
(10) for $\ell_1=1$ imply that
\begin{eqnarray}
H_{2}P=\varepsilon\;,\quad P^{\prime}=-\frac{H_1}{H^{2}_2}\;,\quad
\dot{T}=\lambda T\;,
\end{eqnarray}
where $\varepsilon$ and $\lambda$ are some nonzero constants. One
can easily verify that the first two equations of (43) yield
$\varepsilon^{2}=1$, therefore $\varepsilon=\pm 1$, and that the
first relation implies the second one. The nine equations of (28)
are then as follows for $i=1,2,3$ :
\begin{eqnarray}
\dot{q}_{i}=\frac{m_i}{\lambda T^{2}}\;,\quad
z^{\prime}_i=\frac{m_i}{H^{2}_0}\frac{H_{1}}{P}\;,\quad
z_i^{\prime\prime}-\frac{H_1^{\prime}}{H_1}z_i^{\prime}=\frac{m_i}{H^{2}_0}H^{2}_1\;.
\end{eqnarray}
The first two sets can be easily solved, such that $m_i=2c_iH^{2}_0$
and
\begin{eqnarray}
q_{i}=-c_i\frac{H^{2}_0}{\lambda^{2}T^{2}}+a_i\;,\quad
z_i=c_iH^{2}_2+b_{i} \;,
\end{eqnarray}
where $a_i$ and $b_i$ are new integration constants. These $z_i$
solutions identically satisfy the last relations of (44) without any
extra condition.

By defining $d_i=a_i+b_i$ and
\begin{eqnarray}
v_3=v_3(\varphi)=d_1\cos\varphi +d_2\sin\varphi\;,
\end{eqnarray}
$y_i$ and hence $x,\;y$ functions can be explicitly written from
(15) and (30) as
\begin{eqnarray}
y_i&=&T(c_i S_{-}+ d_i)\;,\quad y=T(S_{-}v_2+v_3)\;,\\
x&=&T(S_{-}v_{2\varphi}+v_{3\varphi})\cos\theta +
T(c_3S_{-}+d_{3})\sin\theta\;,
\end{eqnarray}
where $v_2$ is given by the relation (38) and
\begin{eqnarray}
S_{\pm}=H^{2}_2\pm \frac{H^{2}_0}{\lambda^{2}T^{2}}\;.
\end{eqnarray}
The substitution of $\alpha$ and $\beta$ given by (33) and (34) into
the $\gamma_\varphi$-equation of (2) yields
\begin{eqnarray}
\bar{f}(\varphi)=\frac{T^{2}}{H_0}\partial_t\frac{y}{T}=
2\frac{H_0}{\lambda}v_2 \;.
\end{eqnarray}
Using (38),(47) and (48) in (18) and (34) give the following :
\begin{eqnarray}
\alpha&=&c_4\cos\theta+\sin\theta v\;,\quad
\beta=H_2P\alpha_\theta+2\frac{H_0}{\lambda}H_2v_2\;,\nonumber\\
\gamma&=&2\frac{H_0}{\lambda}H_2(c_3\sin\theta+\cos\theta
v_{2\varphi}
)+H_2Pv_\varphi\;,\nonumber\\
\delta &=& \frac{\lambda}{H_{0}}TH_2\alpha_\theta+T\frac{1}{H_2P}
(S_{+}v_2-v_3)
\;,\\
\epsilon&=&\varepsilon TS_{+}(c_3\sin\theta +\cos\theta
v_{2\varphi})-\varepsilon T(d_3\sin\theta+\cos\theta
v_{3\varphi})+\frac{\lambda}{H_0}TH_2v_{\varphi}\;,\nonumber\\
\mu&=&T(S_{-}v_{2\varphi}+
v_{3\varphi})\sin\theta-T(d_3+c_3S_{-})\cos\theta+c_0TH_2\;,\nonumber
\end{eqnarray}
which determine ten linearly independent KY $2$-forms, one for each
$c_i,\;i=0,1,\ldots,6$ and $d_{j},\;j=1,2,3$. The corresponding KY
2-forms can be written as
\begin{eqnarray}
\omega_{0}&=&TH_2e^{23}\;,\nonumber\\
\omega_{1}&=& 2\frac{H_0}{\lambda}H_2e^{0}\wedge
\Phi_{\varphi}+\varepsilon TS_{+}e^{1}\wedge\Phi-TS_{-}\sin\theta \sin\varphi e^{23}\;,\nonumber\\
\omega_{2}&=& 2\frac{H_0}{\lambda}H_2e^{0}\wedge \Phi-\varepsilon
TS_{+}e^{1}\wedge\Phi_{\varphi}+TS_{-}\sin\theta \cos\varphi e^{23}\;,\nonumber\\
\omega_{3}&=& 2\frac{H_0}{\lambda}H_2\sin\theta e^{03}+\varepsilon
TS_{+}\sin\theta e^{13}
-TS_{-}\cos\theta e^{23}\;,\nonumber\\
\omega_{4}&=& \cos\theta e^{01}-\sin\theta\Psi\wedge e^{2}\;,\\
\omega_{5}&=& \cos\varphi(\sin\theta e^{01}+\cos\theta\Psi\wedge
e^{2})
-\sin\varphi\Psi\wedge e^{3}\;,\nonumber\\
\omega_{6}&=& \sin\varphi(\sin\theta e^{01}+\cos\theta\Psi\wedge
e^{2})
+\cos\varphi\Psi\wedge e^{3}\;,\nonumber\\
\omega_{7}&=&-\varepsilon T\cos\varphi e^{12}+T\sin\varphi(\varepsilon \cos\theta e^{13}-\sin\theta e^{23})\;,\nonumber\\
\omega_{8}&=&-\varepsilon T\sin\varphi
e^{12}-T\cos\varphi(\varepsilon \cos\theta e^{13}-
\sin\theta e^{23})\;,\nonumber\\
\omega_{9}&=&-\varepsilon T\sin\theta e^{13}-T\cos\theta e^{23}
\;,\nonumber
\end{eqnarray}
where the 1-forms $\Phi$ and $\Psi$  are defined, for the sake of
simplicity, as
\begin{eqnarray}
\Phi=\cos\varphi e^{2}-\cos\theta\sin\varphi e^{3}\;,\quad
\Psi=\varepsilon e^{0} +\frac{\lambda}{H_0}T H_2e^{1}\;,
\end{eqnarray}
and $\Phi_\varphi=-(\sin\varphi e^{2}+\cos\theta\cos\varphi e^{3})$.

\section{ KY 2-Forms for $\dot{T}=0\;,\;H^{\prime}_{0}\neq 0\;, \alpha\neq 0$}

In this case, condition (7) requires
\begin{eqnarray}
H^{\prime}_{2}=kH_{0}H_{1}\;,
\end{eqnarray}
such that $k$ is a nonzero constant and, in addition to
$\alpha_\tau=0=\alpha_r$, we immediately have
$\delta_r=0=\delta_\theta$ and $\epsilon_r=0$ from equations (2).
Moreover, the $\epsilon_\varphi$-equation of (2) becomes independent
of the metric characterizing functions, and leads to
$\delta_{\varphi\varphi}+\delta=0$ when substituted into the
$\varphi$-derivative of the second equation of (4). Therefore,
$\delta$ is a harmonic function of $\varphi$ and when this fact is
used in the integration of $\epsilon_\varphi=-\cos\theta \delta$, we
obtain
\begin{eqnarray}
\delta&=&D_1(\tau)\cos\varphi+D_2(\tau)\sin\varphi\;,\nonumber\\
\epsilon&=&\cos\theta\delta_{\varphi}+E(\tau)\sin\theta\;,\\
\mu
&=&-H_{2}P\sin\theta\delta_{\varphi}+E(\tau)H_2P\cos\theta+U(\tau,r)\;,\nonumber
\end{eqnarray}
where we have also made use of the $\mu_\theta$ and
$\mu_\varphi$-equations of (2). The $\epsilon_\theta$-equation then
yields $P^{\prime}=-H_1/H^{2}_2$, which means that $\ell_1=1$ and
$U=H_{2}u(\tau)$.  Therefore, $\alpha$ is again given by (35) and
$\beta$ can be written, from the second equation of (2) and the
first equation of (3), as
$\beta=H_2P\alpha_\theta+H_2B(\tau,\varphi)$. It remains to
determine only the $\tau$ dependence of $\delta,\epsilon$ and $\mu$.

The $\beta_\tau$-equation of (2) and the third equation of (3) give
\begin{eqnarray}
B_\tau=m\delta\;,\quad m_1B=-\delta_\tau\;,
\end{eqnarray}
provided that $m$ and $m_1$ defined by
\begin{eqnarray}
m=\frac{H_{0}^{\prime}}{H_{1}H_{2}}\;,\quad
m_{1}=(\frac{H_{2}}{H_{0}})^{\prime}\frac{H^{2}_{0}}{H_{1}}=kH_{0}^{2}-mH_{2}^{2}\;,
\end{eqnarray}
are new constants, such that $m$ is supposed to be nonzero in this
section. Thus, $\delta$ must satisfy
$\delta_{\tau\tau}+mm_1\delta=0$, and therefore its coefficient
functions can be obtained, depending on the value of $mm_1$, from
\begin{eqnarray}
D_{i\tau\tau}+mm_{1}D_{i}=0\;,\quad i=1,2\;.
\end{eqnarray}
From here on, the discussion proceeds in to ways: (A) $m_1\neq 0$
and (B) $m_1=0$.

\subsection{ KY 2-Forms of the static de Sitter space-time: $m_1\neq 0$ Solutions}

In this case, the $\gamma_\varphi$-equation of (2), the second
equation of (3) and the first equation of (4) specify $\gamma$ as
\begin{eqnarray}
\gamma=H_{2}Pv_\varphi-\frac{H_{2}}{m_1}\cos\theta\delta_{\tau\varphi}+H_2\sin\theta
g(\tau)\;.
\end{eqnarray}
Only three equations remain unused so far; the
$\gamma_\tau,\gamma_\theta$-equations and the equation
$\partial_r(\gamma/H_0)=-H_1\epsilon_\tau/H_0^{2}$ of (3). One can
easily check that the first two of these equations yield
\begin{eqnarray}
g_\tau=mE\;,\quad g=-kE_\tau\;,\quad km_1=1 \;,\quad u_\tau=0\;,
\end{eqnarray}
and the last equation gives nothing new. In accordance with the
previous solutions, we take the constant $u$ as $u=c_0 T$. The first
two equations of (60) also give $E_{\tau\tau}+mm_1E=0$. We can now
write the the complete solutions of the case:
\begin{eqnarray}
\alpha&=&c_4\cos\theta+\sin\theta v\;,\quad
\beta=H_2P\alpha_\theta-\frac{H_2}{m_1}\delta_\tau\;,\nonumber\\
\gamma&=&H_2Pv_\varphi-\frac{H_2}{m_1}(\cos\theta\delta_{\tau\varphi}+\sin\theta
E_{\tau})
\;,\nonumber\\
\delta &=& D_{1}\cos\varphi+D_2\sin\varphi\;,\\
\epsilon&=&\cos\theta \delta_{\varphi}+E\sin\theta\;,\nonumber\\
\mu&=&-H_{2}P\sin\theta\delta_\varphi+
EH_2P\cos\theta+c_0TH_2\;,\nonumber
\end{eqnarray}
where for, say $mm_1=-w_0^{2}<0$, we have
\begin{eqnarray}
D_1&=& a_{1}\cosh w_0\tau+a_{2}\sinh w_0\tau\;,\nonumber\\
D_2&=& a_{3}\cosh w_0\tau+a_{7}\sinh w_0\tau\;,\\
E&=& a_{8}\cosh w_0\tau+a_{9}\sinh w_0\tau\;.\nonumber
\end{eqnarray}
When $mm_1=w^{2}_0>0$, it is enough to replace the hypergeometric
functions of (62) by the corresponding trigonometric functions and
the minus sign by a plus in the expression of $H_0^{2}$. The above
solutions define ten linearly independent KY 2-forms
\begin{eqnarray}
\omega_{0}&=&TH_2e^{23}\;,\nonumber\\
\omega_{1}&=& -\frac{w_0}{m_1}H_2\sinh w_0\tau e^{0}\wedge
\Phi+\cosh w_0\tau (e^1\wedge\Phi+H_{2}P\sin\theta\sin\varphi e^{23}) \;,\nonumber\\
\omega_{2}&=&  -\frac{w_0}{m_1}H_2\cosh w_0\tau e^{0}\wedge
\Phi+\sinh w_0\tau (e^1\wedge\Phi+H_{2}P\sin\theta\sin\varphi e^{23})\;,\nonumber\\
\omega_{3}&=&  \frac{w_0}{m_1}H_2\sinh w_0\tau e^{0}\wedge
\Phi_\varphi-\cosh w_0\tau (e^1\wedge\Phi_{\varphi}+H_{2}P\sin\theta\cos\varphi e^{23})\;,\nonumber\\
\omega_{4}&=& \cos\theta e^{01}-H_2 P\sin\theta e^{02}\;,\\
\omega_{5}&=& \sin\theta\cos\varphi
e^{01}+H_2P(\cos\theta\cos\varphi e^{02}
-\sin\varphi e^{03})\;,\nonumber\\
\omega_{6}&=& \sin\theta\sin\varphi
e^{01}+H_2P(\cos\theta\sin\varphi e^{02}
+\cos\varphi e^{03})\;,\nonumber\\
\omega_{7}&=&\frac{w_0}{m_1}H_2\cosh w_0\tau
e^{0}\wedge\Phi_\varphi- \sinh w_0\tau (e^{1}\wedge
\Phi_\varphi+H_2 P\sin\theta\cos\varphi e^{23})\;,\nonumber\\
\omega_{8}&=&-\frac{w_0}{m_1}H_2\sinh w_0\tau\sin\theta e^{03}+
\cosh w_0\tau (\sin\theta e^{1}+H_2 P\cos\theta e^{2})\wedge e^{3}\;,\nonumber\\
\omega_{9}&=&-\frac{w_0}{m_1}H_2\cosh w_0\tau\sin\theta e^{03}+
\sinh w_0\tau (\sin\theta e^{1}+H_2 P\cos\theta e^{2})\wedge e^{3}
\;,\nonumber
\end{eqnarray}
corresponding, respectively, to $c_i,\;i=0,4,5,6$ and
$a_j,\;j=1,2,3,7,8,9$. Here, $\Phi$ is given by (53) and the metric
coefficient functions are as follows:
\begin{eqnarray}
H_{0}^{2}=m_1^{2}-w^{2}_0H_{2}^{2}\;, \quad
H_1=m_1\frac{H^{\prime}_2}{H_0}\;.
\end{eqnarray}

\subsection{$m_1=0$ Solutions}

When $m_1$ is zero, we have $H_2=m_0H_0$ where $m_0$ is a nonzero
constant and, by virtue of (54) and (57), $k=mm^{2}_0$. Therefore
$k$ and $m$ have the same sign, and (54) and (57) amount to the same
relation. In that case, the equations of (56) are of the forms
$\delta_\tau=0$ and $B_\tau=m\delta$, which imply that
$B=m\tau\delta+C_\varphi(\varphi)$ and
\begin{eqnarray}
\beta=H_2(P\alpha_\theta+m\tau\delta+C_{\varphi})\;,\quad
\delta=b_1\cos\varphi+b_2\sin\varphi\;.
\end{eqnarray}
where $b_i$'s are constants. The $\gamma_\varphi$-equation of (2)
and the second equation of (3) provide us with
\begin{eqnarray}
\gamma=H_2[Pv_\varphi+\cos\theta(m\tau\delta_\varphi-C(\varphi))+G(\tau,\theta)]\;,
\end{eqnarray}
such that $\ell_1=1$. Thus, $\alpha$ is still given by (33) and
$\epsilon,\mu$ are as in equations (55). For the above solutions, we
have $\partial_r(\gamma/H_0)=0$ and the fifth equation of (3)
implies that $\epsilon_\tau=0$, that is, $E$ is independent of time.
On the other hand, the $\gamma_\tau$-equation of (2) and the first
equation of (4) yield
\begin{eqnarray}
G_\tau=mE\sin\theta\;,\quad
C_{\varphi\varphi}=-\sin^{2}\theta\partial_\theta\frac{G}{\sin\theta}\;.\nonumber
\end{eqnarray}
We are finally left with the $\gamma_\theta$-equation, which gives
\begin{eqnarray}
G_\theta=\sin\theta(m\tau\delta_\varphi-C)-\frac{H_2}{H_0}u_\tau\;.\nonumber
\end{eqnarray}
As $G$ is independent of $\varphi$, the only possible solutions of
the last three equations are $G_\theta=-c\sin\theta$ and
$\delta_\varphi=0=u_\tau$ such that $C=c=$constant. Since
$G_{\tau\theta}=G_{\theta\tau}$ implies $E=0$, we can write
$\epsilon=0=\delta$ and $u=c_0T$. Although $G=c\cos\theta$ is a
solution, $c$ does not take part in any component functions. The
results can therefore be written as follows:
\begin{eqnarray}
\alpha&=&c_4\cos\theta+\sin\theta v\;,\quad
\beta=H_2P\alpha_\theta\;,\nonumber\\
\gamma&=&H_2v_\varphi\;,\quad \delta=0=\epsilon\;,\quad \mu=
c_0TH_2\;,\nonumber
\end{eqnarray}
which define four linearly independent KY 2-forms; $\omega_{0}$ and
\begin{eqnarray}
 \omega_{1}&=&\cos\theta e^{01}-mm_0H_{2}\sin\theta
e^{02} \;,\nonumber\\
\omega_{2}&=& -\cos\varphi\omega_{1\theta}-H_{2}\sin\varphi e^{03}\;,\nonumber\\
\omega_{3}&=& -\sin\varphi \omega_{1\theta}+H_{2}\cos\varphi
e^{03}\;.\nonumber
\end{eqnarray}
(The corresponding 1-forms can be found from Section IV-A or V with
$k_1=0$ of the first paper).

\section{KY 2-Forms of The Flat Space-Time}

In this section we take both $\dot{T}$ and $H_0^{\prime}$ to be
zero, which imply
$\beta_\tau=0=\gamma_\tau,\;\delta_r=0=\delta_\theta$ and
$\epsilon_r=0$. Thus, as in the previous section, we have
\begin{eqnarray}
\beta&=&H_2P\alpha_\theta+H_2B(\varphi)\;,\nonumber\\
\delta&=&D_1(\tau)\cos\varphi+D_2(\tau)\sin\varphi\;,\\
\epsilon&=&\cos\theta\delta_{\varphi}+E(\tau)\sin\theta\;,\quad\mu
=H_{2}(P\epsilon_{\theta}+u(\tau))\;,\nonumber
\end{eqnarray}
provided that $P^{\prime}=-H_1/H^{2}_2$ and hence $\alpha$ is given
by (33). On the other hand, the $\beta_r$-equation of (3) gives
\begin{eqnarray}
(H_2P)^{\prime}=0\;,\quad
B=-\frac{H_1}{H_0H^{\prime}_2}\delta_\tau\;.
\end{eqnarray}
The first relation is equivalent to the first integrability
condition of (7), and when it is combined with
$P^{\prime}=-H_1/H^{2}_2$, we obtain $P=\varepsilon/H_2$, that is
$\varepsilon H_2^{\prime}=H_1$ with $\varepsilon=\pm 1$. Five
equations remain untouched so far; the $\gamma_\varphi$-equation of
(2), three equations of (3) involving $\gamma$, and the first
equation of (4).

We first consider the $\gamma_\theta$-equation of (3) by inserting
the $\mu$-solution of (67) into it. The resulting equation implies
that $u$ must be a constant, which we take to be $u=c_0T$, and we
are left with
\begin{eqnarray}
\gamma_\theta=\frac{H^{2}_2P}{H_0}(\sin\theta\delta_{\tau\varphi}-E_\tau\cos\theta)\;.
\end{eqnarray}
As $\gamma$ is independent of $\tau$, $E_\tau$ and $D_{j\tau}$ must
be constants, and therefore $E=a_0\tau+b_0,\; D_1=b_1\tau+b_3$ and
$D_2=b_2\tau+b_4$, where $a_i$ and $b_j$ are constants. Thus, in
terms of
\begin{eqnarray}
z_1(\varphi)=b_1\cos\varphi+b_2\sin\varphi\;,\quad
z_2(\varphi)=b_3\cos\varphi+b_4\sin\varphi\;,
\end{eqnarray}
we can write $\delta=\tau z_1+z_2$, which implies that
$B=-z_1/H_0H_2P$ by virtue of the second relation of (68).
Therefore, (69) can be integrated to obtain the explicit expression
of $\gamma$, presented together with the other solutions below:
\begin{eqnarray}
\alpha&=&c_4\cos\theta+\sin\theta v\;,\quad
\beta=H_2P\alpha_\theta-\frac{1}{H_0P}z_1\;,\nonumber\\
\gamma&=& -\frac{H^{2}_2P}{H_0}(\cos\theta
z_{1\varphi}+a_0\sin\theta)+H_2Pv_{\varphi}\;,\nonumber\\
\delta&=&\tau z_1+z_2\;,\\
\epsilon&=&\tau(\cos\theta z_{1\varphi}+a_0\sin\theta)+\cos\theta
z_{2\varphi}+b_0\sin\theta\;,\nonumber\\
\mu&=&H_{2}(P\epsilon_{\theta}+c_0T)\;.\nonumber
\end{eqnarray}
The last term of $\gamma$ which arises from the $\theta$-integration
mentioned above is specified by the $\beta_\varphi$-equation of (4).
It is straightforward to verify that the remaining three equations
of $\gamma$ are identically satisfied. The corresponding KY 2-forms
are, in addition to $\omega_{0}$, as follows:
\begin{eqnarray}
\omega_{1}&=& -\varepsilon\frac{H_2}{H_0}(\cos\varphi
 e^{02}-\cos\theta\sin\varphi e^{03})+\tau(\cos\varphi e^{12}
-\sin\varphi A\wedge e^{3})\;,\nonumber\\
\omega_{2}&=&  -\varepsilon\frac{H_2}{H_0}(\sin\varphi
e^{02}+\cos\theta\cos\varphi e^{03})+\tau(\sin\varphi e^{12}
+\cos\varphi A\wedge e^{3})\;,\nonumber\\
\omega_{3}&=& \cos\varphi e^{12}-\sin\varphi A\wedge
e^{3}\;,\;\;\;\qquad
\omega_{4}= e^{0}\wedge A\;,\\
\omega_{5}&=& -\cos\varphi e^{0}\wedge A_\theta-\varepsilon\sin\varphi e^{03}\;,\nonumber\\
\omega_{6}&=&  -\sin\varphi e^{0}\wedge
A_\theta+\varepsilon\cos\varphi e^{03}\;, \quad
\omega_{7}=\sin\varphi e^{12}
+\cos\varphi A\wedge e^{3}\;,\nonumber\\
\omega_{8}&=&-\varepsilon\frac{H_2}{H_0}\sin\theta e^{03}-\tau
A_\theta\wedge e^{3}\;,\;\;\quad \omega_{9}=-A_\theta\wedge e^{3}
\;,\nonumber
\end{eqnarray}
where $A=\cos\theta e^{1}-\varepsilon \sin\theta e^{2}$.

\section{Solutions For $\alpha=0$}

In this case the second and fourth equations of (2) give
$\beta_{\theta}=0=\delta_{\theta}$ and
\begin{eqnarray}
\beta=H_{2}B(t,\varphi)\;,\quad\gamma=H_{2}G(t,\theta,\varphi)\;,
\end{eqnarray}
are implied by the first two equations of (3). Fourteen equations
remain to be solved. But when $\alpha$ is set to zero, the
$\gamma_\varphi$ and $\epsilon_\varphi$-equations of (2) are freed
from metric coefficient functions, such that
$\gamma_\varphi=-\cos\theta \beta$ and $\epsilon_\varphi=-\cos\theta
\delta$. When these are combined with two equations of (4), they
considerably ease the investigation by providing general forms of
the solutions. Indeed, differentiating both equations of (4) with
respect to $\varphi$ gives
$\beta_{\varphi\varphi}+\beta=0=\delta_{\varphi\varphi}+\delta$.
Therefore, the general forms of $\beta,\gamma,\delta,\epsilon$ and
$\mu$ can be written, in view of (73), as follows:
\begin{eqnarray}
\beta &=&H_2(B_1(t)\cos\varphi+B_2(t)\sin\varphi)\;,\nonumber\\
\gamma &=&\cos\theta\beta_\varphi+G(t)H_2\sin\theta\;,\nonumber\\
\delta &=&D_1(t,r)\cos\varphi+D_2(t,r)\sin\varphi\;,\\
\epsilon &=&\cos\theta \delta_\varphi+E(t,r)\sin\theta\;,\nonumber\\
\mu&=&(\dot{T}\frac{H_2}{H_0}\beta_\varphi-
\frac{H^{\prime}_2}{H_1}\delta_\varphi)\sin\theta+U(t,r,\theta)\;,\nonumber
\end{eqnarray}
where functions $B_i,D_i, G,E$ and $U$ are to be determined from the
remaining nine equations: the five equations in the first column of
(2) and the last four equations of (3). The second and fourth
relations are obtained by integrations and then by using the results
in equations (4). The last relation of (74) is obtained by first
using $\beta$ and $\delta$ obtained in the last relation of (2) and
then by integrating the resulting equation with respect to
$\varphi$.

When the solutions (74) are substituted into the five equations
appearing in the first column of equations (2), we obtain
\begin{eqnarray}
\dot{B_1}&=&\frac{M}{T}D_1\;,\;\;\quad
\dot{B_2}=\frac{M}{T}D_2\;,\nonumber\\
D_{1r}&=&\dot{T}LB_1\;,\quad
D_{2r}=\dot{T}LB_2\;,\\
\dot{G}&=&\frac{M}{T}E\;,\qquad E_{r}=\dot{T}L G\;,\nonumber\\
U_{\theta}&=&(\dot{T}\frac{H_2^{2}}{H_0}G-\frac{H_2^{\prime}}{H_1}E)\sin\theta\;,\nonumber
\end{eqnarray}
where we have defined the functions
\begin{eqnarray}
M=\frac{H_{0}^{\prime}}{H_{1}H_{2}}\;,\quad
L=\frac{H_{1}H_{2}}{H_{0}}\;.
\end{eqnarray}

\subsection{A General Case}

To be as general as possible, we shall first seek solutions for
which both $\dot{T}$ and $H_0^{\prime}$ can be different from zero.
In such a case, provided that $m$ defined by
\begin{eqnarray}
m =\frac{M^{\prime}}{M^{2}L}\;
\end{eqnarray}
is a constant, from the first four equations of (75) one can easily
obtain
\begin{eqnarray}
B_i=b_iK\;,\quad D_i=b_i\frac{\dot{K}}{K^{m}M}\;;\quad i=1,2\;,
\end{eqnarray}
where $b_1$ and $b_2$ are integration constants, $K=T^{-1/m}$ and
$m$ is supposed to be different from zero. In a similar way, the
fifth and sixth equations of (75) yield
\begin{eqnarray}
G=b_3K\;,\quad E=b_3\frac{\dot{K}}{K^{m}M}\;,\quad
U_\theta=-b_3\frac{\dot{K}R}{K^{m}M}\sin\theta\;,
\end{eqnarray}
where $b_3$ is a constant, and we have defined
\begin{eqnarray}
R=\frac{H_2}{H_1}(m\frac{H_0^{\prime}}{H_0}+\frac{H_2^{\prime}}{H_2})\;.
\end{eqnarray}

In terms of $g=b_1\cos\varphi+b_2\sin\varphi$, the solutions can now
be rewritten as
\begin{eqnarray}
\beta &=&KH_2g\;,\quad\;\; \gamma=KH_2(g_\varphi\cos\theta+b_3\sin\theta)\;,\nonumber\\
\delta &=&\frac{\dot{K}}{K^{m}M}g\;,\quad
\epsilon=\frac{\dot{K}}{K^{m}M}
(g_\varphi\cos\theta+b_3\sin\theta)\;,\\
\mu
&=&-\frac{\dot{K}R}{K^{m}M}(g_\varphi\sin\theta-b_3\cos\theta)+u(t,r)\;,\nonumber
\end{eqnarray}
where the integration of $U_\theta$ done with respect to $\theta$.
There remain the last four equations of (3) that have not been used
so far. Substitution of these solutions into the last four equations
of (3) yield
\begin{eqnarray}
\ddot{K}&=&-m_1K^{2m+1}=-m_2K^{2m+1}\;,\nonumber\\
\partial_t(\frac{u}{T})&=&0=\partial_r(\frac{u}{H_2})
\;,\\
\frac{H_1}{H_2^{2}}&=&-M(\frac{R}{H_2M})^{\prime}\;,\nonumber
\end{eqnarray}
provided that $m_1$ and $m_2$ defined by
\begin{eqnarray}
m_1=M\frac{H_0^{2}}{H_1}(\frac{H_2}{H_0})^{\prime}\;,\quad
m_2=M\frac{H_0}{R}\;,
\end{eqnarray}
are constant. In fact, the first equality of (82) implies that
$m_1=m_2$, and
\begin{eqnarray}
\ddot{K}&=&-m_1K^{2m+1}\;,\quad
R\frac{H_0}{H_1}(\frac{H_2}{H_0})^{\prime}=1\;.
\end{eqnarray}
The equations appearing in the second line of (82) give $u=TH_2$,
and hence  $\mu$  has been completely specified as
\begin{eqnarray}
\mu=-\frac{\dot{K}R}{K^{m}M}(g_\varphi\sin\theta-b_3\cos\theta)+c_0TH_2\;.
\end{eqnarray}
We have obtained five linearly independent KY 2-forms for a family
of space-times characterized by two constants $m$ and $m_1$.

Having determined the coefficient functions of $\omega_{2}$, we now
turn to the conditions which restrict the functions determining the
metric tensor. In addition to two conditions given by (84), we have
two more conditions defining the constants $m$ and $m_1$. The first,
given by (77), can be integrated to yield
\begin{eqnarray}
H_{0}^{\prime}=kH_{0}^{-1/m}H_{1}H_{2}\;
\end{eqnarray}
and hence, $M=kH_0^{m}$. From the first equation of (83) and the
second of (84), we find
\begin{eqnarray}
R=\frac{k}{m_1}H_0^{m+1}\;,\quad
H_2^{\prime}=kH_0^{m-1}H_2^{2}+\frac{m_1}{k}H_1H_0^{-m-1}\;.\nonumber
\end{eqnarray}
Substitutions of these into (73) finally yields
\begin{eqnarray}
m_1H_2(m+\frac{1}{H_1})=H_0^{2}-(\frac{m_1}{k})^{2}H_0^{-2m}\;.
\end{eqnarray}

Although several subclasses of space-times can be identified for
particular values of $m$ and $m_1$, this general consideration will
not be pursued any further. It will suffice to exhibit the
physically important final case.

\subsection{KY $2$-forms of the Robertson-Walker space-time }

The Robertson-Walker space-time is characterized by
\begin{eqnarray}
H_0=1\;,\quad H_2=r\;,\quad H_1^{2}=\frac{1}{1+k_3r^{2}}\;,
\end{eqnarray}
such that $T$ is specified by special cosmological models. Two such
specifications are $T=t^{1/2}$ and $T=t^{2/3}$ which correspond,
respectively, to radiation-dominated and matter-dominated universes.
In this final subsection, we shall present KY 2-forms of this
space-time such that there is no constraint on $T$. It turns out
that such a case is possible only if we take $\alpha=0=\beta=\gamma$
and $H^{\prime}_0=0$. The solutions then are
\begin{eqnarray}
\delta &=&T(c_1\cos\varphi+c_2\sin\varphi)\;,\nonumber\\
\epsilon &=&\cos\theta \delta_\varphi+c_3 T\sin\theta \;,\\
\mu
&=&-H_2P\sin\theta\delta_\varphi+T(c_3H_2P\cos\theta+c_0H_2)\;,\nonumber
\end{eqnarray}
provided that $P^{\prime}=-H_1/H_2^{2}$, which leads to $H_1$ of
(94). The above solutions provide us, in addition to $\omega_0$,
with three linearly independent KY 2-forms:
\begin{eqnarray}
\omega_1 &=&T(\cos\varphi e^{12}-\cos\theta\sin\varphi e^{13})+H_2P\sin\theta\sin\varphi e^{23}\;,\nonumber\\
\omega_2 &=&T(\sin\varphi e^{12}+\cos\theta\cos\varphi e^{13})-H_2P\sin\theta\cos\varphi e^{23}\;,\\
\omega_3 &=&T(\sin\theta e^{13}+H_2P\cos\theta e^{23})\;.\nonumber
\end{eqnarray}

\section{KY 3-Forms}

For a 3-form
\begin{eqnarray}
\omega_{(3)}=\alpha e^{012}+\beta e^{013}+\gamma e^{023}+\delta
e^{123}\;,\nonumber
\end{eqnarray}
the KY-equation gives sixteen equations, five of which have the
following simple forms:
\begin{eqnarray}
\alpha_{t}=0=\beta_{t}\;,\quad \alpha_{r}=0=\beta_{r}\;,\quad
\alpha_{\theta}=0\;.\nonumber
\end{eqnarray}
These imply that $\alpha$ depends only on $\varphi$, and $\beta$ is
a function of $\theta$ and $\varphi$. Seven of the remaining
equations have the following two-term forms:
\begin{eqnarray}
\beta_{\varphi}&=&- \cos\theta \alpha\;,\nonumber\\
\gamma_{t}&=&\frac{H'_0}{T H_1}\delta\;,\quad \gamma_{\theta}
=-\frac{H'_2}{H_1}\beta \;,\; \quad \gamma_{\varphi} =
\frac{H'_2}{H_1}\sin\theta\alpha\;,\\
\delta_{r}&=& \dot{T}\frac{H_1}{H_0}\gamma\;,\quad \delta_{\theta}=-
\dot{T}\frac{H_2}{H_0}\beta\;,\quad \delta_{\varphi} =
\dot{T}\frac{H_2}{H_0}\sin\theta\alpha\;.\nonumber
\end{eqnarray}
The last four equations give three independent equations, which can
be written as:
\begin{eqnarray}
\alpha_{\varphi}=-\sin^{2}\theta\partial_{\theta}\frac{\beta}{\sin\theta}\;,
\quad
\beta_{\theta}=-\frac{H_{2}^{2}}{H_{1}}\partial_{r}\frac{\gamma}{H_{2}}\;,\quad\
\partial_{r}\frac{\gamma}{H_{0}}
=-T^{2}\frac{H_{1}}{H^{2}_{0}}\partial_{t}\frac{\delta}{T}\;.
\end{eqnarray}

Two of the most important integrability conditions for these
equations are as follows:
\begin{eqnarray}
\dot{T}H^{\prime}_{0}\beta = 0\;,\quad
[\rho(r)-\varrho(t)]\beta=0\;,
\end{eqnarray}
where $\rho$ and $\varrho$ are given by (7). The first follows from
$\gamma_{t\theta}=\gamma_{\theta t}$ and
$\delta_{r\theta}=\delta_{\theta r}$, and the second from
$\delta_{t\theta}=\delta_{\theta t}$. There are also identical
conditions with $\beta$ replaced by $\alpha$ that can be checked
from $\gamma_{t\varphi}=\gamma_{\varphi t}\;,
\delta_{r\varphi}=\delta_{\varphi r}$ and
$\delta_{t\varphi}=\delta_{\varphi t}$. But noting that $\beta=0$
implies $\alpha=0$, we see that the above conditions include the
second (see also relations (94)). There are also some other
conditions which should be considered in investigating the cases
implied by the above conditions. Therefore, we shall present the
general solutions in two classes: (A) $\beta\neq 0$ and (B)
$\beta=0$.

The essence of spherical symmetry seems to be encoded in the first
equations of (91) and (92), for they do not depend on the metric
coefficient functions. We are thus able to start with their general
solutions
\begin{eqnarray}
\alpha= a_1\cos\varphi+a_2 \sin\varphi\;,\quad \beta =
\cos\theta\alpha_\varphi+a_3\sin\theta\;,
\end{eqnarray}
which can easily be verified. Here, $a_i$ are integration constants.

\subsection{Solutions for $\beta\neq0$}

For the fulfilment of conditions (93) in the case of nonzero
$\beta$, two sets of conditions must be distinguished:
\begin{eqnarray}
{\bf (i)}: \;\;\quad \dot{T} &=& 0\;,\quad
H^{\prime}_{2}=kH_{0}H_{1}\;,\nonumber\\
{\bf (ii)}: \quad H_{0}^{\prime} &=& 0\;,\quad
\dot{T}^{2}-T\ddot{T}=-\ell=\frac{H_{0}^{2}}{H_{1}H_{2}}(H_2P)^{\prime}\;,\nonumber
\end{eqnarray}
Here $k$ and $\ell$ are constants such that $k\neq 0$. The
well-known maximal symmetric Minkowski and the static form of de
Sitter space-times, each having five independent KY $3$-forms are
obtained among the ${\bf (i)}$ solutions as special cases. On the
other hand, four time dependent forms plus the most well-known form
of de Sitter and Robertson-Walker space-times, emerge in the second
case. We should emphasize the fact that the former space-time is
obtained without any restriction on $T$, which is obtained by taking
$H_0^{\prime}$ zero and by starting in such a way that the last two
conditions of {\bf (ii)} are avoided.

\subsubsection{KY 3-Forms for the Minkowski and Static Form of de Sitter Space-time}

In the {\bf (i)} case we have, in addition to $\delta=f(\tau)$, the
following five equations:
\begin{eqnarray}
\gamma_{\tau} &=&\frac{H'_0}{H_1}f\;,\quad\qquad \gamma_{\theta}
=-kH_0\beta\;,\;\quad \gamma_{\varphi} =
kH_{0}\sin\theta\alpha\;,\nonumber\\
&&\\
f_{\tau}&=&-\frac{H^{2}_{0}}{H_{1}}\partial_{r}\frac{\gamma}{H_{0}}\;,\;\;
\beta_{\theta}=-\frac{H_{2}^{2}}{H_{1}}\partial_{r}\frac{\gamma}{H_{2}}\;.\nonumber
\end{eqnarray}
where $\tau=t/T$. Provided that $m$ is a nonzero constant such that
\begin{eqnarray}
H_{2}H_{0}^{\prime}-H_{0}H_{2}^{\prime}=mH_{1}\;,
\end{eqnarray}
the last two equations of (95) imply that $m\gamma$ must be equal to
$H_{2}f_{\tau}-H_{0}\beta_{\theta}$.  The first and second equations
of (95) in this case require $km=-1$ and $f_{\tau\tau}+m_{1}f = 0$,
where the constant $m_1$ is defined by
\begin{eqnarray}
H_{0}^{\prime}&=&km_{1}H_{1}H_{2}\;.
\end{eqnarray}
The third equation of (95) is then identically satisfied.

As an alternative approach one can first integrate the
$\gamma_{\theta}$-equation with respect to $\theta$, and then use it
in the $\gamma_{\varphi}$-equation to find $\gamma
=kH_{0}\beta_{\theta}+G(\tau,r)$. The remaining three equations then
give the same solution. As a result, under the ${\bf (i)}$
conditions the general forms of the coefficient functions for KY
$3$-form are, in addition to that given by (94), as follows:
\begin{eqnarray}
\gamma =-k(f_{\tau}H_{2}-H_{0}\beta_{\theta})\;,\quad \delta =
f(\tau)\;.
\end{eqnarray}

Depending on the value of $m_{1}$, one can easily write the explicit
form of $f$. For $m_1=0$ we have, in terms of integration constants
$a,b$,
\begin{eqnarray}
f=a+b\tau\;,\quad H_{2}^{\prime}=kH_0 H_1\;.
\end{eqnarray}
and $H^{2}_{0}=k^{-2}$ from equations (96) and (97). For nonzero
$m_1$, we have
\begin{eqnarray}
H^{2}_{0}=k_0+m_1H_{2}^{2}\;,\quad H_0 H_1=k_0
kH_{2}^{\prime}\;,\quad k_0 k^{2}=1\;
\end{eqnarray}
and $f$ is as follows ($k_0,b_1$ and $b_2$ are integration
constants):
\begin{eqnarray}
f=\left\{
\begin{array}{cc}
b_1\cos\omega_0\tau+b_2\sin\omega_0\tau\;; & \quad m_1=\omega_0^{2}>0\;, \\
b_1\cosh\omega_0\tau+b_2\sinh\omega_0\tau\;, &\quad
m_1=-\omega_0^{2}<0\;,
\end{array}
\right.
\end{eqnarray}
In any case, we have five linearly independent $3$-forms.

For $T=1,\; H_0=1=H_1$ and $H_2=r$ we get, from the equations (94),
(97) and (99), 3-forms of Minkowski space-time:
\begin{eqnarray}
\omega_1 &=& w_1 +\sin\theta \sin\varphi e^{023}\;,\quad \omega_2 =
w_2
-\sin\theta \cos\varphi e^{023}\;,\nonumber\\
\omega_3 &=& \sin\theta e^{013}+\cos\theta e^{023}
\;,\\
\omega_4 &=& e^{123}\;,\quad\omega_5 =-r e^{023}+\tau e^{123}\;,
\nonumber
\end{eqnarray}
where the $3$-forms $w_1$ and $w_2$ are defined, for brevity, as
\begin{eqnarray}
w_1 = \cos\varphi e^{012}-\cos\theta \sin\varphi e^{013}\;,\quad w_2
= \sin\varphi e^{012}+\cos\theta \cos\varphi e^{013} \;.
\end{eqnarray}
Note that $k=1$ (hence $m=-1$) and $m_1=0$ for this case. On the
other hand, for the values
\begin{eqnarray}
T=1\;,\quad H_2=r\;,\quad k_0=1=k \;
\end{eqnarray}
we obtain, by virtue of equations (94), (97) and (100), the KY
$3$-forms
\begin{eqnarray}
\omega_1 &=& w_1 +H_0\sin\theta \sin\varphi e^{023}\;,\quad \omega_2
= w_2
-H_0\sin\theta \cos\varphi e^{023}\;,\nonumber\\
\omega_3 &=& \sin\theta e^{013}+H_0\cos\theta e^{023} \;,\quad
\omega_4 =r\sinh\tau e^{023}+\cosh\tau e^{123}\;,\\
\omega_5 &=&-r\cosh\tau e^{023}+\sinh\tau e^{123}\; \nonumber
\end{eqnarray}
for the static form of de Sitter space-time, specified also by
$H_1^{2}=1+m_1r^{2}$ and $H_0H_1=1$.

\subsubsection{Solutions for Four Time-Dependent Forms of de Sitter Space-time}

From here on, we consider the ${\bf (ii)}$ conditions. The condition
$H_{0}^{\prime}=0$ gives $\gamma_{t}=0$ and leaves us with seven
equations to be solved. It is easy to integrate the
$\gamma_\theta$-equation of (91) and then use it in the
$\gamma_\varphi$-equation to find $\gamma=H_2P\beta_\theta+G(r)$. By
substituting this solution into the $\beta_\theta$-equation, we find
$G=cH_2$ and hence, $\gamma=H_2P\beta_\theta+cH_2$, provided that
$P^{\prime}=-H_{1}/H_{2}^{2}$. Here, $c$ is an integration constant.
The $\delta_\theta$-equation can also be integrated with respect to
$\theta$, and then by substituting the solution into
$\delta_\varphi$-equation, we find
\begin{eqnarray}
\delta=\dot{T}\frac{H_2}{H_0}\beta_\theta+D(t,r)\;.
\end{eqnarray}

The following two equations of (91) and (92) remain to be solved:
\begin{eqnarray}
\delta_{r} =\dot{T}\frac{H_{1}}{H_{0}}\gamma\;,\;\quad
\frac{H_{0}}{H_{1}}\gamma_{r} =
-T^{2}\partial_{t}\frac{\delta}{T}\;.\nonumber
\end{eqnarray}
By substituting the above $\gamma$ solution and (106) into these
equations, we obtain
\begin{eqnarray}
D_{r}=c\dot{T}\frac{H_{1}H_2}{H_{0}}\;,\quad
T^{2}\partial_{t}\frac{D}{T}=-cH_{0}H_2P\;,
\end{eqnarray}
provided that
\begin{eqnarray}
(H_2P)^{\prime}=-\ell \frac{H_{1}H_2}{H_{0}^{2}} \;,
\end{eqnarray}
which is just one of the integrability conditions of {\bf (ii)}. Now
it is not difficult to see that for nonzero values of $\ell$, the
general solution for $D$ is $D=-(c/\ell)\dot{T}H_{0}H_2P+c_0T$,
where $c_0$ is another integration constant. We can now collate the
solutions of the case as follows:
\begin{eqnarray}
\alpha &=& a_1\cos\varphi+a_2 \sin\varphi\;,\quad\beta=
\cos\theta\alpha_\varphi+a_3\sin\theta
\;,\nonumber\\
\gamma &=&H_2P\beta_\theta+cH_2\;,\\
\delta
&=&\dot{T}\frac{H_2}{H_0}\beta_\theta-\frac{c}{\ell}\dot{T}H_{0}H_2P+c_0T\;.\nonumber
\end{eqnarray}
Depending on the values of $\ell$ and other integration constants
arising when integrating the equation
$T^{2}\partial_t(\dot{T}/T)=\ell$, four different time regimes were
presented in Part I (see relations (72) in I). The above solutions
provide five independent KY $3$-forms:
\begin{eqnarray}
\omega_1 &=& \cos\varphi e^{012}-\cos\theta\sin\varphi
e^{013}+H_2\sin\theta\sin\varphi B\;,\nonumber\\
\omega_2 &=& \sin\varphi e^{012}+\cos\theta\cos\varphi
e^{013}-H_2\sin\theta\cos\varphi B\;,\nonumber\\
\omega_3 &=& \sin\theta e^{013}+H_2\cos\theta B \;,\\
\omega_4 &=&H_2(e^{023}-\frac{H_0}{\ell}\dot{T}Pe^{123})\;,\nonumber\\
\omega_5 &=&T e^{123}\;, \nonumber
\end{eqnarray}
where we have defined the 3-form
$B=Pe^{023}+H^{-1}_0\dot{T}e^{123}$.

\subsubsection{KY 3-Forms of de Sitter Space-Time with Exponential Time Dependence }

When $\ell=0$, we have $\dot{T}=\lambda T$ and $H_{2}^{\prime}=m_0
H_1$ from (108). In such a case, the most general solution of (107)
turns out to be
\begin{eqnarray}
D=c\frac{\lambda T}{2m_0H_0}[H_{2}^{2}+(\frac{m_0H_0}{\lambda
T})^{2}]+c_1\frac{1}{T}\;,\nonumber
\end{eqnarray}
where $c_1$ is an integration constant. The solutions are then given
by (109), with $\delta$ replaced by
$\delta=\dot{T}(H_2/H_0)\beta_\theta+D$. From $P^{\prime}=-H_1/H_2$,
it follows that $m_0^{2}=1$, that is, $m_0=\varepsilon=\pm1$ and we
again obtain five linearly independent $3$-forms for de Sitter
space-time. The first three forms are the same as those given in
(110), and the last two are as follows:
\begin{eqnarray}
\omega_4 =H_2e^{023}+\frac{\lambda
T}{2m_0H_0}[H_{2}^{2}+(\frac{m_0H_0}{\lambda T})^{2}]e^{123}\;,\quad
 \omega_5 =\frac{1}{T} e^{123}\;.
\end{eqnarray}

\subsubsection{KY 3-form of the Robertson-Walker Space-Time}

For the solutions obtained so far under ${\bf (ii)}$ conditions, $T$
is restricted as a special function of time. It turns out (see also
the next section) that the only possible solution in which there are
no constraints on $T$ is
\begin{eqnarray}
\omega=Te^{123}\;,
\end{eqnarray}
provided that $H_0^{\prime}=0$. In particular, there  is only one KY
3-form for the Robertson-Walker space-time.

\subsection{Solutions for $\beta=0$}

For $\beta=0$, the equations (91) and (92) imply that
$\alpha=0,\;\gamma=H_{2}f(t)$ and $\delta=D(t,r)$, and the following
three equations
\begin{eqnarray}
T\dot{f}&=&\frac{H_{0}^{\prime}}{H_{1}H_2}D\;,\nonumber\\
\dot{T}f&=&\frac{H_{0}}{H_{1}H_2}D_r\;,\\
\frac{T^{2}}{f}\partial_t\frac{D}{T}&=&
-\frac{H_{0}^{2}}{H_{1}}(\frac{H_2}{H_0})^{\prime}\;,\nonumber
\end{eqnarray}
determine $f$ and $D$. If $H_{0}^{\prime}$ is zero, then $f$ is a
constant, say $c$, and we get  $\gamma=cH_{2}$ and
$\delta=-(c/\ell)\dot{T}H_0H_2P+c_0T$, which are special cases of
solutions (109). Moreover, the special case $c=0$ produces the
solution (112) for the Robertson-Walker space-time.

For nonzero $H_0^{\prime}$, taking $D$ from the first equation of
(113) and substituting it into the other two equations lead us to
two separate equations. Each side of these equations must be
constant such that
\begin{eqnarray}
\frac{H_{0}}{H_{1}H_{2}}(\frac{H_{1}H_2}{H_{0}^{\prime}})^{\prime}=m_1\;,\quad
\frac{H_{0}^{2}H_{0}^{\prime}}{H^{2}_{1}H_2}(\frac{H_2}{H_0})^{\prime}=m_2\;.
\end{eqnarray}
This leaves us with two simple equations for $f$:
\begin{eqnarray}
\frac{\dot{T}f}{T\dot{f}} =m_1\;,\quad T^{2}\frac{\ddot{f}}{f}=
-m_2\;.
\end{eqnarray}
For $m_1=0$, we have $\dot{T}=0$ and
\begin{eqnarray}
f_{\tau\tau}+m_2f=0\;,\quad H_{0}^{\prime}=m_{3}H_{1}H_{2}\;,\quad
D=m_{3}^{-1}f_{\tau}.
\end{eqnarray}
where $m_3$ is a nonzero constant. These provide two independent
$3$-forms:
\begin{eqnarray}
\omega_1 &=& \tau H_2e^{023}+\frac{1}{m_3}e^{123}\;,\quad \omega_2 =
H_2e^{023}\;,\nonumber
\end{eqnarray}
for $m_2=0$ and
\begin{eqnarray}
\omega_1 &=& H_2\cosh w_0\tau
e^{023}+\frac{w_0}{m_3}\sinh w_0\tau e^{123}\;,\nonumber\\
\omega_2 &=& H_2\sinh w_0\tau e^{023}+\frac{w_0}{m_3}\cosh w_0\tau
e^{123}\;,\nonumber
\end{eqnarray}
for $m_2=w_0^{2}>0$. For nonzero values of $m_1$, the first equation
of (115) can be easily integrated to yield $f=c_1 T^{1/m_1}$, and by
substituting this into the second equation, we get
$\ddot{K}=-m_2K^{1-2m_1}$, where $K=T^{1/m_1}$. As long as this last
condition and that given by (114) are satisfied, the general
solutions which define only one KY 3-form are as follows:
\begin{eqnarray}
\gamma=c_{1}H_{2}T^{1/m_1}\;,\quad
\delta=\frac{c_{1}}{m_1}\dot{T}T^{1/m_1}\frac{H_1H_2}{H_{0}^{\prime}}\;.\nonumber
\end{eqnarray}

\section{Conclusion}

By directly starting from the KY-equation, we have developed a
constructive method which makes it possible to generate all KY forms
for a large class of spherically symmetric space-times in a unified
and exhaustive way. Our results for the well-known spherically
symmetric space-times are quantitatively summarized in Table I of
the first paper and their KY two and three forms are computed in
this second paper. In particular, we have found an exactly solvable
nonlinear time equation for de Sitter type space-times which enables
us to generate all of their KY forms in a unified manner. We have
also reported solutions in some detail for sufficiently symmetric
new cases which fall within the considered class of metrics.

Our results can be used to reach decisive, or at least conclusive
statements in analyzing the algebraic structures of KY-forms
\cite{Gibbons,Kastor1,Cariglia}, in specifying of the symmetry
algebra and related conserved quantities of the Dirac as well as
other equations in spherically symmetric curved backgrounds
\cite{Benn-Charlton,Benn-Kress1,Benn-Kress2,Benn-Kress3,Cebeci}.
Finally, as an application, we indicate an approach for calculating
Killing tensors and associated first integrals for the considered
class of space-times \cite{Benn4}. As has been mentioned before, to
each KY (p+1)-form $\omega$, there corresponds an associated Killing
tensor $K$ that can be defined by
$K(X,Y)=g_{p}(i_X\omega,i_Y\omega)$, where $g_{p}$ is the compatible
metric in the space of $p$-forms induced by the space-time metric
$g$. Then  $i_{\dot{\gamma}}\omega$ is parallel-transported along
the affine-parameterized geodesic $\gamma$ with tangent field
$\dot{\gamma}$, and $K(\dot{\gamma},\dot{\gamma})$ is the associated
quadratic first integral. The first statement follows from the fact
that the covariant and interior derivatives with respect to the same
geodesic tangent field commute and the second statement follows from
the fact that the cyclicly permuted sum of $\nabla_{X}K(Y,Z)$
vanishes. In particular, Killing tensor fields and associated first
integrals for the space-times given in the Table I of I can be
computed and used in investigating some integrability problems. Our
study on the symmetries of the Dirac equation and related matter, is
in progress, and soon will be reported elsewhere \cite{ozumav2}.

\begin{acknowledgments}
This work was supported in part by the Scientific and Technical Research Council of
Turkey (T\"{U}B\.{I}TAK).
\end{acknowledgments}

\end{document}